\documentstyle[12pt]{article}
\newcommand{\be}{\begin{equation}}
\newcommand{\ee}{\end{equation}}

\newcommand{\bea}{\begin{eqnarray}}
\newcommand{\eea}{\end{eqnarray}}
\newcommand{\ba}{\begin{array}}
\newcommand{\ea}{\end{array}}
\newcommand{\beqa}{\begin{eqnarray}}
\newcommand{\eeqa}{\end{eqnarray}}

\newcommand{\NP}[1]{Nucl. Phys.\ {\bf #1}\ }
\newcommand{\PL}[1]{Phys. Lett.\ {\bf #1}\ }

\newcommand{\PR}[1]{Phys. Rev.\ {\bf #1}\ }
\newcommand{\PRL}[1]{Phys. Rev. Lett.\ {\bf #1}\ }

\newcommand{\h}{{1\over2}}
\newcommand{\qu}{{1\over4}}
\newcommand{\del}{\partial}

\newcommand{\Tr}{{\rm Tr}}

\newcommand{\D}{\delta}
\newcommand{\DE}{\Delta}

\newcommand{\ka}{\kappa}

\newcommand{\ie}{{\it i.e. }}

\newcommand{\ssu}{$SU(2)_L\times SU(2)_R\times U(1)_{B-L}\,$}

\newcommand{\sur}{$SU(2)_R$}
\newcommand{\matr}{\left( \begin{array}}
\newcommand{\ematr}{\end{array} \right)}

\newcommand{\lsim}
{{\;\raise0.3ex\hbox{$<$\kern-0.75em\raise-1.1ex\hbox{$\sim$}}\;}}
\newcommand{\gsim}
{{\;\raise0.3ex\hbox{$>$\kern-0.75em\raise-1.1ex\hbox{$\sim$}}\;}}

\begin{document}

\begin{titlepage}

\mbox{}\vspace*{-1cm}\hspace*{8cm}\makebox[7cm][r]{\large  HU-SEFT R
1994-14} \mbox{}\vspace*{-0cm}\hspace*{9cm}\makebox[7cm][r]{\phantom
{HU-SEFT R 1994-14}} \vspace*{0.5cm}
 \hspace*{10.9cm} \makebox[4cm]{(hep-ph/9410342)} \vfill

\Large

\begin{center}
{\bf The Higgs sector of a supersymmetric left-right  model}

\bigskip
\normalsize
{K. Huitu$^a$ and J. Maalampi$^b$ }\\
[15pt]
{\it $^a$Research Institute for High Energy Physics, University of Helsinki
\\$^b$Department of Theoretical Physics,
University of Helsinki}

{October 1994}

\vfill

\normalsize

{\bf\normalsize \bf Abstract}

 \end{center}

\normalsize
\noindent
We study the symmetry breaking  sector of a supersymmetric
left-right model based on the gauge group \ssu .
The explicit mass matrices of neutral, singly charged and doubly
charged scalars are constructed.
In the minimum the $R$-parity is found to be spontaneously broken.
An experimentally interesting feature of the model is that one of the
doubly charged scalars is possibly light enough to be seen in the next
linear collider.

\end{titlepage}

\newpage

\setcounter{page}{2}

\noindent
{\it 1. Introduction.}
Supersymmetry is often invoked to take care of the quadratic
divergences occurring in the scalar sector of spontaneously broken gauge
theories.
The Higgs sector of supersymmetric models has proven to be interesting
in view of future colliders,
since the mass of the lightest neutral scalar has typically
a relatively low upper limit.
In the framework of the minimal supersymmetric standard model (MSSM)
the tree level mass of the neutral Higgs is bound from above by the $Z$ boson
mass.
However, radiative corrections to the scalar masses can be large
\cite{ERZ}.

In this paper, we will study the Higgs sector of a supersymmetric left-right
model
(SLRM) based on the gauge group {\ssu }.
We construct the mass matrices of the physical Higgs scalars
and determine the mass spectra for a representative choice of free
parameters.

The motivation for the left-right models is mainly the see-saw
mechanism \cite{seesaw} by which one can generate light masses for the
left-handed neutrinos and large masses for the right-handed ones.
The left-right models are especially interesting, if
the experiments on solar \cite{sun}
and atmospheric \cite{atmos} neutrinos
continue to show deviation from the standard model, as well as the
existence of the hot dark matter component \cite{COBE} explaining some features
of the power spectrum of density fluctuations of the Universe persists.
All these results seem to indicate that neutrinos indeed have a
small mass.

To achieve the see-saw mechanism, the {\ssu } symmetry
has to be broken by scalar triplets of \sur .
The special feature of the model is that the triplets contain among
others also doubly charged scalars.
The phenomenology related to the supersymmetric partners of these
scalars has been recently studied in \cite{HMR}.
The gauge symmetry breaking of the supersymmetric left-right
model was also studied in \cite{CP,KM}.
In \cite{KM}  it
was argued that the parity is violated only if the $R$-parity
is broken.
Here we will see in an explicit construction that the tree-level
masses of the pseudoscalars and
doubly charged scalars can be physical only if the $R$-parity
is spontaneously broken.
\vfill

\pagebreak
\noindent
{\it 2. The scalar potential of a supersymmetric left-right model
and the gauge symmetry breaking.}
The most general potential \cite{DGKO}
of the standard left-right model
is complicated due to the numerous possible combinations
of the fields.
In the supersymmetric version, the Higgs couplings are much more
constrained, since the quartic interactions
are completely determined by the gauge couplings.

The superpotential of the model is given by

\bea
W & = & h_ {\phi Q} \widehat Q_{L}^{T}i\tau_2 \widehat \phi  \widehat Q_{R}^c
+ h_{ \chi Q} \widehat Q_{L}^{T} i\tau_2\widehat
\chi  \widehat Q_{R}^c \nonumber \\
&&+h_ {\phi L} \widehat L_{L}^{T}i\tau_2 \widehat \phi  \widehat L_{R}^c
+h_{ \chi L} \widehat L_{L}^{T} \widehat i\tau_2 \widehat\chi
 \widehat L_{R}^c
+h_{\DE} \widehat L_{R}^{cT} i\tau_2
{\widehat \DE }  \widehat L_{R}^c \nonumber\\
&& + \mu_1 {\rm Tr} (i\tau_2 \widehat \phi^T i\tau_2 \widehat \chi )
+\mu_2  \Tr (\widehat \DE \widehat \D ) ,\label{pot}
\eea

\noindent
where $\widehat Q_{L (R)}$ denote the left (right) handed quark superfield
doublets and similarly for the leptons $\widehat L_{L (R)}$.
The triplet and the bidoublet Higgs superfields of {\ssu } are given by

\bea
\label{higgses}
&&\widehat\DE =\matr{cc}\widehat\DE^-/\sqrt{2} & \widehat\DE^{0}\\
 \widehat\DE^{--}&-\widehat\DE^{-}/\sqrt{2} \ematr
\sim ({\bf 1,3,}-2),\nonumber\\
&&\widehat\delta =\matr{cc}\widehat\delta^{+}/\sqrt{2}& \widehat\delta^{++}\\
 \widehat\delta^{0} &-\widehat\delta^{+}/\sqrt{2} \ematr
\sim ({\bf 1,3,}2),
\,\,\,\,\,\,\,\,\,\,\,\, \nonumber\\
&& \widehat\phi =\matr{cc}\widehat\phi_1^0&
\widehat\phi_1^+\\\widehat\phi_2^-&\widehat\phi_2^0
\ematr \sim ({\bf 2,2,}0),
\,\,\,\,\,\,\,\,\,\,\,\,
\widehat\chi =\matr{cc}\widehat\chi_1^0&
\widehat\chi_1^+\\\widehat\chi_2^-&\widehat\chi_2^0
\ematr  \sim ({\bf 2,2,}0).
\eea

\noindent
Corresponding to each scalar multiplet with non-zero $U(1)$
quantum number,
one has to include another multiplet with an opposite $U(1)$
quantum number in order to avoid chiral anomalies for the fermionic
superpartners.
Also another bidoublet Higgs superfield is added to get a nontrivial
Kobayashi-Maskawa matrix.

The left-right model contains often also the $SU(2)_L$ triplet so as to make
the
 Lagrangian fully symmetric under the  $L\leftrightarrow R$ transformation.
Phenomenologically this is not, however, necessary, since the left
triplet is not needed for the symmetry breaking or the
see-saw mechanism.
As our purpose is to study a minimal phenomenologically viable model,
we have not included the $SU(2)_L$ triplets in our considerations.

To explore the symmetry breaking and to work out the mass spectra of  the Higgs
sector, one has to consider the so-called F-terms and D-terms, as well as
the possible soft supersymmetry breaking terms.
Those of such terms which contain electrically neutral scalars are given by

\bea
\label{scapot}
&&V_D=\nonumber\\
&& \sum_i \left\{
\h g_L^2 \left|\h \Tr \phi^\dagger \tau_i \phi +\h \Tr \chi^\dagger
\tau_i \chi \right. +\h \tilde L_L^\dagger \tau_i \tilde L_L\right|^2
 \nonumber \\
&&+\left. \h g_R^2 \left|\h \Tr \phi^\dagger \tau_i \phi +
\h \Tr \chi^\dagger \tau_i \chi
 + \Tr\Delta ^\dagger \tau_i \Delta +\Tr\delta ^\dagger \tau_i \delta
+\h \tilde L_R^{*\dagger} \tau_i \tilde L_R^*\right|^2  \right\}
\nonumber \\
&& +\h g_{B-L}^2 \left|-\Tr \Delta ^\dagger  \Delta
+\Tr \delta ^\dagger  \delta
-\h\tilde L_L^{\dagger} \tilde L_L
+\h\tilde L_R^{*\dagger} \tilde L_R^*
\right|^2 , \nonumber\\
&& \nonumber \\
&&V_{soft}+V_F =\nonumber\\
&& m_1^2\Tr |\phi|^2 + m_2^2\Tr |\chi|^2 -
(m_{\phi \chi}^2
\Tr (i\tau_2 \phi^T i\tau_2) \chi +h.c.) \nonumber\\
&&+m_{3 }^2 |\Delta |^2 + m_{4 }^2|\delta |^2
-(m_{\Delta\delta}^2\Tr \Delta\delta +h.c.)+m_5^2 |\tilde L_R^*|^2
+m_6^2 |\tilde L_L|^2 \nonumber\\
&&+ (\tilde L_R^T (A_\phi\phi +A_\chi\chi )\tilde L_L
+A_\DE \tilde L_R^{*T} i\tau_2 \DE \tilde L_R^* +h.c. )
\nonumber\\
&&+|h_{\phi L} i\tau_2 \phi \tilde L_R^* +h_{\chi L} i\tau_2 \chi
\tilde L_R^*|^2 +
|h_{\phi L} \tilde L_L^T i\tau_2\phi + h_{\chi L} \tilde L_L^T i\tau_2\chi
+2h_\DE \tilde L_R^{*T}i\tau_2 \DE |^2 \nonumber \\
&&
+\Tr|h_\DE \tilde L_R^* (\tilde L_R^{*T} i\tau_2) +\mu_2 \delta |^2
+|-h_{\chi L } i\tau_2 \tilde L_L \tilde L^{*T}_R +\mu_1 i\tau_2 \phi
i\tau_2 |^2
\nonumber \\ &&
+|-h_{\phi L } i\tau_2 \tilde L_L \tilde L^{*T}_R +\mu_1 i\tau_2 \chi
i\tau_2 |^2,
\eea

\noindent
where the scalar fields are denoted by the same symbols than the
superfields, except for the hat.
The soft supersymmetry breaking parameters are the soft trilinear
couplings $A_i$ and the soft masses which are contained in $m_i$.
The $m_i$'s are defined so that the coefficients of the F-terms
of similar form are included.

We do not make an effort to minimize the full scalar potential, but
instead find a region in the parameter space for which
the scalar fields in the minimum have
vacuum expectation values given as follows:

\bea
\langle\phi \rangle=
\left( {\begin{array}{cc} \ka_1 & 0\\ 0 & 0 \end{array}}
\right),&&
\: \langle\chi \rangle=
\left( {\begin{array}{cc} 0 & 0\\ 0 & \ka_2 \end{array}}
\right),\:  \label{phivev}\nonumber\\
\langle\Delta\rangle=\left( {\begin{array}{cc} 0 &  v_\DE\\0 & 0 \end{array}}
\right)
 , &&\:   \langle\delta \rangle=\left(
{\begin{array}{cc} 0 & 0\\ v_\delta & 0 \end{array}} \right)\nonumber\\
\langle\tilde\nu_R \rangle=\left( {\begin{array}{c} 0 \\ \sigma_R \end{array}}
\right),
&& \: \langle\tilde\nu_L \rangle=\left( {\begin{array}{c} \sigma_L \\ 0
\end{array}}
\right).
\label{vevs}
\eea

\noindent
Also $\phi_{2   }^0$ and $\chi_{1 }^0$ in Eq. (\ref{higgses})
could get
vev's without breaking the electric charge, but as explained
in \cite{HMR} on phenomenological grounds $\langle\phi_{2}^0\rangle, \,
\langle\chi_1^0 \rangle$
are much smaller than  $\langle\phi_{1}^0\rangle,\,\langle\chi_ 2^0\rangle$,
since the mixing between the charged gauge bosons has to be tiny.

The Yukawa coupling $h_{\chi L}$ is proportional to the neutrino
Dirac mass $m_D$.
The light neutrino mass in the see-saw mechanism is
given by $\sim m_D^2/m_M$, where $m_M=h_{\Delta}v_{\Delta}$ is  the Majorana
mass. The value of the triplet Higgs vev $v_{\Delta}$ gives the scale of the
SU(2)$_R$ breaking and is, according to the lower bounds of the heavy W- and
Z-boson masses, in the range $v_{\Delta}\gsim$  1 TeV (see below).
Therefore the
magnitude of $h_{\chi L}$ is still quite inaccurately determined given the
present upper limits for the light neutrino masses. On the other hand the
Yukawa
coupling $h_{\phi L}$ is  proportional to the electron mass and we will neglect
it.

The $m_i$'s, $i=1,...,5$ appearing in (\ref{scapot}), can be eliminated by
using
the minimization conditions $\del V/\del\ka_2=\del V/\del\ka_1=
\del V/\del v_\DE=\del V/\del v_\delta=\del V/\del \sigma_R =
\del V/\del \sigma_L =0$.
For simplicity we will assume in the following that only one of the vev's
of the neutral scalar
leptons is non-negligible, namely $\sigma_R$.
This will fix the  parameter $\mu_1$ as
$\mu_1= 2 h_\DE \ka_2 v_\DE/\ka_1$.
Imposing the conditions for minimum, the potential
is given as

\bea
V|_{min}&=&-\frac 18 g_L^2 (\ka_1^2-\ka_2^2)^2
-\h g_{B-L}^2(v_\DE^2-v_\delta^2-\h \sigma_R^2)^2 \nonumber \\
&&-\h g_R^2 (v_\DE^2-v_\delta^2-\h \sigma_R^2-\h (\ka_1^2-\ka_2^2))^2
+ \sigma_R^2(- 4h_\DE^2v_\DE^2 -
h_\DE^2\sigma_R^2  -h_{\chi L}^2 \ka_2^2 \nonumber\\
&&+  A_\DE v_\DE
+h_\DE\mu_2 v_\delta ) .
\eea

\noindent
In the true minimum this has to be negative.
For the first three terms this is obvious, but the last
term is constrained by the requirement of
negativity.

In the breakdown of the gauge symmetries down to the $U(1)_{em}$
the charged gauge bosons and two of the neutral ones become massive.
The masses are found by diagonalizing the corresponding mass
matrices and they are given by

\bea
m_{Z_1}^2&=& \h (g_L^2+g'^2 )(\ka_1^2+\ka_2^2)
+O\left(\frac {(\ka_1^2+\ka_2^2)^2}{v_\DE^2+v_\delta^2
+\qu \sigma_R^2}
\right)\label{z1} ,\\
m_{W_L}^2&=& \h g_L^2(\ka_1^2+\ka_2^2)  ,\\
m_{Z_2}^2&=&2g_R^2\left[\qu \frac{g'^2}{g_{B-L}^2}(\ka_1^2+\ka_2^2) +
\frac{g_{B-L}^2}{g'^2}(v_\DE^2+v_\delta^2+\qu \sigma_R^2)\right]
+O\left(\frac {(\ka_1^2+\ka_2^2)^2}{v_\DE^2+v_\delta^2
+\qu \sigma_R^2} \right) ,\nonumber\\
&&\\
m_{W_R}^2&=& g_R^2(v_\DE^2+v_\delta^2
+\h \sigma_R^2+ \h (\ka_1^2+\ka_2^2)) .
\label{w2}
\eea

\noindent
Here it is denoted $g'=g_Rg_{B-L}/\sqrt{g_R^2+g_{B-L}^2}$.

In the following we will assume that the left- and right-couplings are
equal to the standard model $SU(2)$ gauge coupling $g$, $g_L=g_R=g$.
This assumption also determines the value of the third coupling, $g'$
by (\ref{z1}) up to a correction factor.
Then the $\rho $-parameter of the electroweak interactions
is found to be given by

\bea
&&\rho^{-1} =\frac {M_{Z_1}^2\cos^2\theta_W}{M_{W_L}^2} \nonumber \\
&& = 1-\frac 1{4}
\frac {g'^4}{g_{B-L}^4}
\frac {\ka_1^2+\ka_2^2}
{v_\DE^2+v_\delta^2 +\qu \sigma_R^2}
+O\left(\frac {(\ka_1^2+\ka_2^2)^2}{(v_\DE^2+v_\delta^2 +\qu \sigma_R^2)^2}
\right),
\eea

\noindent
where the angle $\theta_W$ is given by $\tan\theta_W= g'/g$.
The effect of the new scalars getting vev's is to increase the value of $\rho
$.
As expected, the value of the $\rho $ parameter approaches its
standard model value as $v_\DE$, $v_\delta$, or $\sigma_R$ get large values.
The experimental limits for the $\rho $-parameter are given by
$\rho=0.998\pm 0.0086$ \cite{AC}.
With the above mentioned assumptions on the gauge couplings, the
experimental bound for $\rho $ constrains the ratio of the vacuum
expectation values as follows

\be
\frac {\ka_1^2+\ka_2^2} {v_\DE^2+v_\delta^2 +\qu \sigma_R^2} < 0.053 .
\ee

\noindent
Since the sum of the vacuum expectation values $\ka_ {1,2}^2$
is determined from the measured mass of
$W_L$ \cite{PDG},
the charged heavy gauge boson mass
$m_{W_R} \gsim 500$ GeV.
\vfill

\pagebreak
\noindent
{\it 3. Spontaneous breaking of R-parity.}
In the minimum the masses of all the scalars in the Higgs sector must be
positive.
The scalar masses can be found by using the scalar potential, Eq.
(\ref{scapot}),
and the vacuum expectation values given in Eq. (\ref{vevs}).
Before studying the ensuing mass spectrum numerically,
let us consider the $R$-parity, $R=(-1)^{3(B-L)+2s}$.
As discussed in \cite{FIQ}, the $R$-parity, which is +1 for ordinary
particles and -1 for their supersymmetric counterparts, is automatically
conserved in Lagrangian in this type of models, but it
may be broken spontaneously if $\langle\tilde\nu \rangle\not= 0$.
In the neutral mass matrix the pseudoscalar and scalar components do
not mix.
In the case of conserved $R$-parity, \ie $\langle\tilde\nu_{R,L} \rangle=0$,
the pseudoscalar mass matrix
is given by four
two by two blocks, see Eqs. (\ref{na2}) and (\ref{na6}) in the Appendix.
One of the blocks contains the sneutrinos and we need not consider
it here.
Two of the blocks contain the Goldstone bosons which make
two of the neutral gauge bosons massive.
The physical pseudoscalar particles have the masses

\bea
m_{A_1}^2&=& m_{\phi \chi}^2\left(\frac {\ka_1}{\ka_2} + \frac {\ka_2}{\ka_1}
\right) ,
\nonumber \\
m_{A_2}^2&=& m_{\Delta \delta }^2\left( \frac {v_\delta }{v_\DE} +
\frac {v_\DE }{v_\delta}
\right) ,
 \nonumber \\
m_{A_{3,4}}&=& \h \left\{  \right. m_{A_1}^2 \pm \left[  \right. m_{A_1}^4+
4(m_{W_R}^2\cos 2\gamma  -m_{W_L}^2\cos 2\beta )^2
\nonumber \\
&&  -4(m_{W_R}^2\cos 2\gamma  -m_{W_L}^2\cos 2\beta )
m_{A_1}^2 \cos 2\beta \left.\left. \right] ^{1/2} \right\} ,
\eea

\noindent
where we have defined

\be
\tan\beta =\frac {\ka_2}{\ka_1} , \;\; \tan\delta
=\frac {v_\delta }{ v_\DE }  ,\,\,
{\rm and }\;\;
\tan^2\gamma = \frac{ v_\delta ^2 + \h \ka_1^2}{ v_\DE ^2 + \h \ka_2^2}.
\ee

\noindent
On the other hand the masses of the doubly charged scalars are given by

\be
m_{H^{++}_{1,2}}^2=\h\left\{ m_{A_2}^2\pm \sqrt{m_{A_2}^4
+8 m_{W_R}^2\cos 2\gamma
[m_{A_2}^2 \cos 2 \delta + 2m_{W_R}^2 \cos 2 \gamma ]} \right\}.
\ee

\noindent
It is easily seen that trying to make both pseudoscalar and doubly charged
masses positive one ends in contradiction.
Necessarily at least one of the $\langle\tilde\nu \rangle\not= 0$.
\vfill

\pagebreak
\noindent
{\it 4. Scalar mass spectrum.}
Let us now investigate the mass spectrum of the scalars predicted by the
model.
The physical Higgses in the model consist of eight neutral scalars,
six pseudoscalars, six singly charged scalars, and two doubly
charged scalars.
The corresponding mass matrices are given in the Appendix in
Eqs. (\ref{nh2}) - (\ref{npp}).
Throughout the numerical calculations it is assumed that $g_L=g_R =g$
and the mass of $W_R$ is taken to be 1 TeV.
In this case the value of the $\rho $-parameter increases from its standard
model value by $\DE\rho=\rho -1\sim 0.0017$.

The mass parameters containing the soft breaking terms are chosen to be 1 TeV.
As usual in the susy models, it turns out that the experimentally
most interesting, lightest scalar masses, both neutral and
singly or doubly charged, depend only very slightly on
the soft masses.
E.g. changing $m_{\phi\chi }$ and $m_{\DE\delta }$ from 1 TeV
to 2 TeV increases the light masses by less than 5 GeV.
We have also assumed that $m_5=m_6$.

The parameters of the model are constrained by the requirement that
all the masses remain real in the allowed range.
Reality of the pseudoscalar masses leads to an upper limit of
$h_{\chi L}$ which varies between about 0.2 and 0.4.
To obtain more insight into the parameters of the mass matrices, we
study first a specific limit with $h_{\chi L} = 0 = \mu_2$ and
D-terms, which are negligible.
In this limit the doubly charged masses are real if
$A_\DE > 4h_\DE^2 v_\DE $.
On the other hand one finds that the neutral scalar masses can be real
only if $A_\DE $ is in the range

\be
 \frac{h_\DE^2}{2v_\DE} \left[\sigma_R^2 + 8 v_\DE^2 -
\sqrt{\sigma_R^4+16 \sigma_R^2 v_\DE^2} \right] < A_\DE <
 \frac{h_\DE^2}{2v_\DE} \left[\sigma_R^2 + 8 v_\DE^2 +
\sqrt{\sigma_R^4+16 \sigma_R^2 v_\DE^2} \right].
\ee

\noindent
With these restrictions the masses of pseudoscalars and singly charged
scalars are real.
The allowed range disappears in the limit $h_\DE \rightarrow 0$.
In a more general situation with $h_{\chi L}=0.1$ and $g=0.65$
the coupling $A_\DE$ as a function of $h_\DE$ is plotted
in Fig. 1
for two values of the supersymmetric mass parameter $\mu_2$
($\mu_2=100$ GeV and 1 TeV).
The masses of the doubly charged scalars are positive above the lower
curve, whereas the masses of the neutral scalars are positive below the upper
curve.
The relatively narrow range between the curves is thus the allowed range of
$h_\DE$ and $A_\DE$.
For very small values of $h_\DE$ this range almost disappears.
Here we have taken $\tan\beta=50$, $v_\delta /\sigma_R =1.5$, and
$v_\DE/v_\delta =1.05$.
The curves depend only very slightly on $\tan\beta $.

In Table 1 and 2 we show a typical spectra of the physical scalars and give
their
compositions for $\tan\beta=1.5$ and $\tan\beta=50$.
We also give compositions of the unphysical Goldstone bosons needed
to get massive gauge bosons.
The lightest neutral scalar contains mostly bidoublet  fields $\phi_1^{0r}$
and $\chi_2^{0r}$ whereas the
heaviest is mostly a combination of the triplet fields $\DE^{0r}$,
$\delta^{0r}$, and the right-sneutrino $\tilde \nu_R^r $.
Similar compositions are found
for the lightest and heaviest pseudoscalars and the singly
charged scalars.
Varying the supersymmetric mass parameter $\mu_2$ between 100 GeV and
1 TeV has an effect of a few GeV on the mass of the lightest neutral
Higgs, whereas the effect on the lightest singly
charged Higgs is negligible.
The lightest neutral scalar resembles closely
the lightest neutral scalar of MSSM.
Consequently, the radiative corrections to $m_{H^0_1}$ are large
due to the top and stop loops \cite{ERZ}.
One may expect also large radiative corrections to the masses
of those neutral scalars, which contain large portions of $\DE^0 $,
due to the heavy right-handed neutrino contributions.
However, the scalars containing $\DE^0$ tend to be heavy already
in the tree level.

{}From the experimental point of view the most interesting situation
arises when the doubly charged scalar is light.
It is easy to detect (the signature is two leptons of same charge)
and it is typical for the left-right model.
The variation of the $H^{++}$ masses as a function of the allowed
$A_\DE$ values is plotted in Fig. 2 for $h_\DE=0.3,\ldots ,0.8$.
The mass increases fast from zero as $A_\DE$ increases.
The solid curve corresponds again to $\mu_2 =1$ TeV and the dashed one
to $\mu_2 =100$ GeV.
Increasing $\mu_2 $ changes the place of the curve in $A_\DE $ axis, but
the behaviour of the mass with increasing $A_\DE $ is very similar.
The maximum value of $m_{H^{++}}$ is typically between 200 GeV and
400 GeV.

The relation between the mass of the doubly charged Higgs and the possibly
non-diagonal Yukawa coupling to the leptons
have been studied in \cite{mlS,LP}. The most stringent constraint comes from
the
upper limit for the decay $\mu\to\overline eee$ \cite{mlS}:

\be
h_{\DE,e\mu}h_{\DE,ee}<4.7\times 10^{-11}{\rm GeV}^{-2}\times m_{H^{++}} ^2.
\ee
{}From the Bhabha-scattering
cross section at SLAC and DESY the following bound  for the $h_{\DE ,ee}$
coupling was established:

\be
h_{\DE ,ee}^2 < 9.7 \times 10^{-6} {\rm GeV }^{-2}\times m_{H^{++}} ^2.
\ee

\noindent
For $h_{\DE ,ee}$=0.6 the mass of the doubly charged boson
$m_{H^{++}} \gsim 200$ GeV.
For the coupling $h_{\DE ,\mu\mu}$ the muonium transformation to antimuonium
converts into a limit $h_{\DE ,ee} h_{\DE ,\mu\mu } < 5.8 \times 10^{-5}
{\rm GeV}^{-2}\times m_{H^{++}}^2 $  \cite{mlS}.

\vspace{.5in}
\noindent
{\it 5. Summary.}
The physical Higgses in the supersymmetric {\ssu } electroweak
model we have studied consist of eight neutral scalars,
six pseudoscalars, six singly charged scalars, and two doubly
charged scalars.
We have shown that non-vanishing vacuum expectation values of
bidoublet and triplet Higgses, sufficent for the breaking of the gauge
symmetry, do not produce a physical mass spectrum, but also at least
one of the sneutrinos should acquire a vev.
Hence R-parity is necessarily broken in the model.
With a viable choice of the various mass parameters we find that there are a
few relatively light scalars or pseudoscalars with masses of the order of
100-200 GeV.
These include a phenomenologically quite interesting
doubly charged scalar with a mass about 200 GeV.
The majority of the scalar particles are in the mass range above 1 TeV.

\setcounter{equation}{0}
\renewcommand{\theequation}{A\arabic{equation}}
\appendix
\vspace{.5in}
\noindent
{\bf Appendix.}

In this Appendix we give the mass matrices of neutral, singly charged and
doubly charged scalars.
In the mass matrices we denote

\bea
&&\ka_1^2-\ka_2^2=\kappa_{dif}^2 ,  \\
&&v_\DE^2-v_\delta^2-\h \sigma_R^2-\h (\ka_1^2- \ka_2^2) = \rho _{dif}^2.
\eea

\noindent
In the neutral mass matrix the pseudoscalar and scalar components do
not mix.
We denote the
pseudoscalars by a superscript $i$ and scalars by $r$.
The mass matrix of the neutral scalar sector consists of one $2\times 2$
and one $6\times 6$ block:

\be
\label{nh2}
M^2_{\phi_{2  }^{0r},\chi_{1 }^{0r}}=
\matr{cc} m_{\phi \chi}^2\frac{\ka_2}{\ka_1}-\h  g_L^2 \kappa_{dif}^2
+ g_R^2 \rho_{dif}^2 & -m_{\phi \chi}^2 \\-m_{\phi \chi}^2 & m_{\phi \chi}^2
\frac{\ka_1}{\ka_2}
+\h g_L ^2\kappa_{dif}^2 -g_R^2 \rho_{dif}^2 -h_{\chi L}^2 \sigma_R^2
\ematr,
\ee

\noindent
and $(M_{\tilde\nu_L^{r} ,\,\tilde\nu_R^{*r},\, \phi_{1   }^{0r},\,
\chi_{2 }^{0r},\,\Delta^{0r} , \,\D^{0r} }^2)_{ij}=
(M^{0,r(6)})_{ij} $ where the non-zero terms are given by

\bea
\label{nh6}
(M^{0,r(6)})_{\tilde\nu_L\tilde\nu_L} &=& m_6^2 + h_{\chi L}^2 (\ka_2^2+
\sigma_R^2)
+\qu (g_L^2+g_{B-L}^2) \ka_{dif}^2+\h g_{B-L}^2 \rho_{dif}^2, \nonumber \\
(M^{0,r(6)})_{\tilde\nu_L\tilde\nu_R^*}&=&  -2h_{\chi L} h_\DE\ka_2 v_\DE
+h_{\chi L}\mu_1\ka_1,
\nonumber \\
(M^{0,r(6)})_{\tilde\nu_L\phi_{1  }^{0} }&=& h_{\chi L}\mu_1 \sigma_R,
\nonumber\\
(M^{0,r(6)})_{\tilde\nu_L\chi_{2 }^{0}}&=& -2h_{\chi L} h_\DE \sigma_R v_\DE,
\nonumber \\
(M^{0,r(6)})_{\tilde\nu_L\DE^0}&=& -2h_{\chi L} h_\DE \sigma_R\ka_2, \nonumber
\\
(M^{0,r(6)})_{\tilde\nu_R^* \tilde\nu_R^*}&=&
4 h_\DE^2 \sigma_R^2  +\h (g_{B-L}^2+g_R^2)\sigma_R^2, \nonumber \\
(M^{0,r(6)})_{\tilde\nu_R^*\phi_{1   }^{0}}&=& \h g_R^2\ka_1\sigma_R,
\nonumber \\
(M^{0,r(6)})_{\tilde\nu_R^*\chi_{2 }^{0}}&=&
(2 h_{\chi L}^2 -\h g_R^2) \sigma_R\ka_2,\nonumber \\
(M^{0,r(6)})_{\tilde\nu_R^*\DE^0}&=& (8 h_\DE^2 -g_R^2-g_{B-L}^2)v_\DE
\sigma_R -2A_\DE \sigma_R,\nonumber \\
(M^{0,r(6)})_{\tilde\nu_R^*\delta^0 }&=& (g_R^2+g_{B-L}^2)v_\delta \sigma_R
-2h_\DE\mu_2 \sigma_R, \nonumber \\
(M^{0,r(6)})_{\phi_{1 }^{0}\phi_{1 }^{0}}&=& m_{\phi \chi}^2
\frac{\ka_2}{\ka_1}
+\h (g_R^2+g_L^2)\ka_1^2, \nonumber \\
(M^{0,r(6)})_{\phi_{1 }^{0}\chi_{2 }^{0}}&=& -m_{\phi \chi}^2
-\h (g_R^2+g_L^2)\ka_1\ka_2, \nonumber \\
(M^{0,r(6)})_{\phi_{1 }^{0}\DE^0}&=&-g_R^2\ka_1 v_\DE,\nonumber \\
(M^{0,r(6)})_{\phi_{1 }^{0}\delta^0}&=&g_R^2\ka_1 v_\delta,  \nonumber \\
(M^{0,r(6)})_{\chi_{2 }^{0}\chi_{2 }^{0}}&=&  m_{\phi
\chi}^2\frac{\ka_1}{\ka_2}
+\h (g_R^2+g_L^2)\ka_2^2, \nonumber \\
(M^{0,r(6)})_{\chi_{2 }^{0}\DE^0} &=& g_R^2 \ka_2v_\DE, \nonumber \\
(M^{0,r(6)})_{\chi_{2 }^{0} \delta^0} &=& -g_R^2\ka_2 v_\delta, \nonumber \\
(M^{0,r(6)})_{\DE^{0}\DE^{0}}&=& m_{\DE\delta}^2\frac {v_\delta}{v_\DE}
+A_\DE \frac{\sigma_R^2}{v_\DE}+2(g_R^2+g_{B-L}^2)v_\DE^2, \nonumber \\
(M^{0,r(6)})_{\DE^{0}\delta^{0}}&=& -m^2_{\DE\delta}
-2(g_R^2+g_{B_L}^2)v_\delta v_\DE,\nonumber \\
(M^{0,r(6)})_{\delta^{0}\delta^{0}}&=&m_{\DE\delta}^2\frac{v_\DE}{v_\delta}
+h_\DE\mu_2\frac{\sigma_R^2}{v_\delta} +2(g_R^2+g_{B-L}^2)v_\delta^2.
\eea

\vfill
\pagebreak

The pseudoscalar mass matrix is of block diagonal form of one  two
by two block and one six by six block:

\be
\label{na2}
M_{\phi_{2 }^{0i},\,\chi_{1 }^{0i}}^2=
\matr{cc} m_{\phi  \chi}^2 \frac{\ka_2 }{\ka_1 }-\h  g_L^2 \ka_{dif}^2+
g_R^2\rho_{dif}^2
&m_{\phi  \chi}^2 \\
m_{\phi  \chi}^2 &
m_{\phi  \chi}^2 \frac{\ka_1 }{\ka_2 }+\h  g_L^2 \ka_{dif}^2- g_R^2
\rho_{dif}^2 -h_{\chi L}^2\sigma_R^2
\ematr
\ee

\noindent
and
$(M_{\tilde\nu_L^{i} ,\,\tilde\nu_R^{*i},\, \phi_{1  }^{0i},\,\chi_{2 }^{0i},
\,\Delta^{0i} , \,\D^{0i} }^2)_{ij}=
(M^{0,i(6)})_{ij}, $
with the non-zero terms given by

\bea
\label{na6}
(M^{0,i(6)})_{\tilde\nu_L\tilde\nu_L} &=&m_{6}^2 + h_{\chi L}^2 (\ka_2+
\sigma_R^2)+
\qu (g_L^2+g_{B-L}^2) \ka_{dif}^2 +\h g_{B-L}^2 \rho_{dif}^2,  \nonumber \\
(M^{0,i(6)})_{\tilde\nu_L\tilde\nu_R^*}&=& -2h_{\chi L}h_\DE \ka_2 v_\DE
-h_{\chi L}\mu_1\ka_1, \nonumber \\
(M^{0,i(6)})_{\tilde\nu_L\phi_{1 }^{0} }&=& h_{\chi L} \mu_1 \sigma_R,
\nonumber\\
(M^{0,i(6)})_{\tilde\nu_L\chi_{2 }^{0}}&=& 2 h_{\chi L} h_\DE \sigma_R v_\DE,
\nonumber \\
(M^{0,i(6)})_{\tilde\nu_L\DE^0}&=& -2h_{\chi L} h_\DE \sigma_R\ka_2, \nonumber
\\
(M^{0,i(6)})_{\tilde\nu_R^* \tilde\nu_R^*}&=& 4 A_\DE v_\DE +
4 h_\DE\mu_2 v_\delta, \nonumber \\
(M^{0,i(6)})_{\tilde\nu_R^*\DE^0}&=& 2 A_\DE \sigma_R, \nonumber \\
(M^{0,i(6)})_{\tilde\nu_R^*\delta^0 }&=& -2h_\DE\mu_2\sigma_R, \nonumber \\
(M^{0,i(6)})_{\phi_{1 }^{0}\phi_{1 }^{0}}&=& m_{\phi  \chi}^2 \frac{\ka_2}
{\ka_1}, \nonumber \\
(M^{0,i(6)})_{\phi_{1 }^{0}\chi_{2 }^{0}}&=& m_{\phi  \chi}^2, \nonumber \\
(M^{0,i(6)})_{\chi_{2 }^{0}\chi_{2 }^{0}}&=&m_{\phi \chi}^2 \frac{\ka_1}
{\ka_2}, \nonumber \\
(M^{0,i(6)})_{\DE^{0}\DE^{0}}&=&m_{\DE\delta}^2\frac{v_\delta}{v_\DE}
+ \sigma_R^2\frac{A_\DE}{v_\DE}, \nonumber \\
(M^{0,i(6)})_{\DE^{0}\delta^{0}}&=&m_{\DE\delta}^2,\nonumber \\
(M^{0,i(6)})_{\delta^{0}\delta^{0}}&=&m_{\DE\delta}^2\frac{v_\DE}{v_\delta}
+h_\DE\mu_2\frac{\sigma_R^2}{v_\delta}.
\eea

The singly charged mass matrix consists of one three by three block and one
five by five block as follows:

\vfill
\pagebreak

\bea
&&
M_{\tilde e_L^*,\phi_{2 }^+,\chi_{1 }^+}^2= \nonumber\\
&&\matr{ccc}
m_6^2+h_{\chi L}^2\sigma_R^2-\qu (g_L^2 -g_{B-L}^2)\ka_{dif}^2
+\h g_{B-L}^2 \rho_{dif}^2 & \mu_1h_{\chi L} \sigma_R &
2 h_{\chi L} h_\DE \sigma_R v_\DE \\
 \mu_1h_{\chi L} \sigma_R &
m_{\phi \chi}^2 \frac{\ka_2}{\ka_1} +\h g_L^2\ka_2^2
& m_{\phi \chi}^2 +\h g_L^2 \ka_1\ka_2\\
2 h_{\chi L} h_\DE \sigma_R v_\DE &
 m_{\phi \chi}^2 +\h g_L^2 \ka_1\ka_2
& m_{\phi \chi}^2 \frac{\ka_1}{\ka_2} +\h g_L^2\ka_1^2
\ematr \nonumber \\
\eea

\noindent
and $(M_{\tilde e_R^*,\phi_{1 }^+,\chi_{2 }^+,\DE^+,\delta^+}^2)_{ij}
\equiv (M^{+(5)})_{ij}$ with
\bea
(M^{+(5)})_{\tilde e_R^*\tilde e_R} &=& -4h_\DE^2v_\DE^2
+2h_\DE\mu_2 v_\delta +2 A_\DE v_\DE +g_R^2\rho_{dif}^2
+\h g_R^2 \sigma_R^2 - h_{\chi L}^2 \ka_2^2, \nonumber \\
(M^{+(5)})_{\tilde e_R^*\phi_{1 }^-} &=& -\h g_R^2\sigma_R\ka_1, \nonumber \\
(M^{+(5)})_{\tilde e_R^*\chi_{2 }^-} &=& (h_{\chi L}^2 -\h g_R^2 )
\sigma_R\ka_2,
 \nonumber \\
(M^{+(5)})_{\tilde e_R^*\DE^-} &=& -\sqrt{2}(\h g_R^2v_\DE +A_\DE
-2 h_\DE^2 v_\DE )\sigma_R, \nonumber \\
(M^{+(5)})_{\tilde e_R^*\delta^-} &=& \sqrt{2}(\h g_R^2v_\delta -h_\DE\mu_2)
\sigma_R,
 \nonumber \\
(M^{+(5)})_{\phi_{1 }^+\phi_{1 }^-} &=& m_{\phi \chi}^2\frac{\ka_2}{\ka_1}
 +g_R^2 \rho_{dif}^2 +\h g_R^2\ka_1^2,\nonumber \\
(M^{+(5)})_{\phi_{1 }^+\chi_{2 }^-} &=& m_{\phi \chi}^2 +\h g_R^2\ka_1\ka_2,
\nonumber \\
(M^{+(5)})_{\phi_{1 }^+\DE^-} &=& \frac1{\sqrt{2}} g_R^2 \ka_1 v_\DE,
\nonumber \\
(M^{+(5)})_{\phi_{1 }^+\delta^-} &=& -\frac1{\sqrt{2}} g_R^2 \ka_1 v_\delta,
\nonumber \\
(M^{+(5)})_{\chi_{2 }^+\chi_{2 }^-} &=& m_{\phi \chi}^2\frac {\ka_1}{\ka_2}
 -g_R^2 \rho_{dif}^2 +\h g_R^2\ka_2^2 - h_{\chi L}^2 \sigma_R^2, \nonumber \\
(M^{+(5)})_{\chi_{2 }^+\DE^-} &=& \frac{1}{\sqrt{2}} g_R^2
\ka_2 v_\DE,\nonumber \\
(M^{+(5)})_{\chi_{2 }^+\delta^-} &=& -\frac{1}{\sqrt{2}} g_R^2
 \ka_2 v_\delta, \nonumber \\
(M^{+(5)})_{\DE^+\DE^-} &=& m^2_{\DE\delta}\frac {v_\delta}{v_\DE}
-2 h_\DE^2 \sigma_R^2 +\frac {A_\DE }{v_\DE }\sigma_R^2
+\h g_R^2\ka_{dif}^2
+g_R^2(v_\delta^2 +\h \sigma_R^2 ), \nonumber \\
(M^{+(5)})_{\DE^+\delta^-} &=& -m_{\DE\delta}^2-g_R^2v_\DE v_\delta, \nonumber
\\
(M^{+(5)})_{\delta^+\delta^-} &=& m^2_{\DE\delta}\frac {v_\DE}{v_\delta}
+h_\DE\mu_2 \frac{\sigma_R^2}{v_\delta} -\h g_R^2\ka_{dif}^2
+g_R^2(v_\DE^2 -\h \sigma_R^2 ).
\eea

\vfill
\pagebreak

The doubly charged mass matrix can be read from the scalar potential to be

\bea
\label{npp}
&&M^2_{\DE^{++},\delta^{++}}=\nonumber \\
&&{\scriptstyle \left( \begin{array}{cc} {m_{\DE\delta}^2\frac{v_\delta
}{v_\DE}
-\sigma_R^2 (4h_\DE^2 -\frac{A_\DE }{v_\DE }) -2g_R^2 \rho_{dif}^2}
& {-m_{\DE\delta}^2} \\ {-m_{\DE\delta}^2} &
{m_{\DE\delta}^2\frac{v_\DE}{v_\delta }
+h_\DE\mu_2 \frac{\sigma_R^2}{v_\delta } + 2g_R^2\rho _{dif}^2 }
\end{array}\right) }.
\eea

\vspace{2cm}
\large
\noindent Acknowledgements
\normalsize
\vspace{0.5cm}

The work has been supported by
the  Academy of Finland.

\begin{table}
\begin{tabular}{|l|r|l|} \hline
particle & mass [GeV] & composition \\ \hline
$H^0_1$ & 39 & $-0.01\tilde\nu_R^{r}-0.56 \phi^{0r}_{1 }-0.83\chi^{0r}_{2 }
-0.01\delta^{0r} $\\
$H^0_2$ & 115 & $0.57 \phi^{0r}_{2  }+0.82\chi^{0r}_{1 } $\\
$H^0_3$ & 514 & $0.41 \tilde\nu_R^{r}
-0.01 \phi_1^{0r}-0.01 \chi_2^{0r}+0.64\DE^{0r}
+0.65\delta^{0r} $\\
$H^0_4$ & 862 & $0.90 \tilde\nu_R^{r}
-0.16\DE^{0r}-0.42\delta^{0r} $\\
$H^0_5$ & 1409 & $0.78\tilde\nu_L^{r}-0.52\phi^{0r}_{1  }
+0.35\chi^{0r}_{2}-0.02\DE^{0r}+0.01\delta^{0r} $\\
$H^0_6$ & 1466 & $0.82 \phi^{0r}_{2  }-0.57\chi^{0r}_{1 } $\\
$H^0_7$ & 1515 & $0.63\tilde\nu_L^{r}+0.65\phi^{0r}_{1   }
-0.43\chi^{0r}_{2 }+0.02\DE^{0r}-0.02\delta^{0r} $\\
$H^0_8$ & 2252 & $0.16 \tilde\nu_R^{r}+0.02\phi^{0r}_{1  }
-0.03\chi^{0r}_{2 }-0.75\DE^{0r}+0.64\delta^{0r} $\\
$A_1$ & 115 & $-0.57 \phi^{0i}_{2   }+0.82\chi^{0i}_{1 } $\\
$A_2$ & 1396 & $0.04\tilde\nu_L^{i}+0.25 \tilde\nu_R^{i}-
0.02\phi^{0i}_{1 }-0.01\chi^{0i}_{2 }-
0.62\DE^{0i} -0.74\delta^{0i}$\\
$A_3$ & 1409 & $0.77\tilde\nu_L^{i}-0.01 \tilde\nu_R^{i}-0.53\phi^{0i}_{1 }
-0.35\chi^{0i}_{2 }+0.03\DE^{0i} +0.03\delta^{0i}$\\
$A_4$ & 1466 & $0.82 \phi^{0i}_{2  }+0.57\chi^{0i}_{1 } $\\
$A_5$ & 1514 & $0.64\tilde\nu_L^{i}+0.64 \phi^{0i}_{1  }
+0.43\chi^{0i}_{2 }$\\
$A_6$ & 2869 & $-0.01\tilde\nu_L^{i}+0.94 \tilde\nu_R^{i}
+0.33\DE^{0i} +0.04\delta^{0i}$\\
$H^\pm_1$ & 121 & $-0.02\tilde e_R^{\pm}-0.57 \phi^{\pm}_{1  }
+0.82\chi^{\pm}_{2 }-0.06\DE^\pm -0.05\delta^\pm $\\
$H^\pm_2$ & 1329 & $0.78\tilde e_R^{\pm}-0.05 \phi^{\pm}_{1  }
-0.03\chi^{\pm}_{2 }+0.20\DE^\pm -0.58\delta^\pm $\\
$H^\pm_3$ & 1409 & $-0.77\tilde e_L^{\pm}+0.53 \phi^{\pm}_{2 }
+0.35\chi^{\pm}_{1 }$\\
$H^\pm_4$ & 1467 & $+0.07\tilde e_R^{\pm}+0.82 \phi^{\pm}_{1  }
+0.57\chi^{\pm}_{2 }-0.02\DE^\pm -0.01\delta^\pm $\\
$H^\pm_5$ & 1516 & $0.63\tilde e_L^{\pm}+0.64 \phi^{\pm}_{2  }
+0.43\chi^{\pm}_{1 }$\\
$H^\pm_6$ & 1928 & $-0.53\tilde e_R^{\pm}+0.03 \phi^{\pm}_{1  }
+0.03\chi^{\pm}_{2 }+0.69\DE^\pm -0.48\delta^\pm $\\
$H^{\pm\pm}_1$ & 224 & $0.68\DE^{\pm\pm} +0.73\delta^{\pm\pm} $\\
$H^{\pm\pm}_2$ & 1433 & $0.73\DE^{\pm\pm} -0.68\delta^{\pm\pm} $\\
Goldstone$^0_1$& 0 &$-0.11 \tilde\nu_R^{i}-0.48 \phi^{0i}_{1  }
+0.71\chi^{0i}_{2 }+0.36\DE^{0i} -0.34\delta^{0i}$\\
Goldstone$^0_2$ & 0 &$-0.14 \tilde\nu_R^{i}+0.44 \phi^{0i}_{1  }
-0.66\chi^{0i}_{2 }+0.43\DE^{0i} -0.41\delta^{0i}$\\
Goldstone$^\pm_1$ & 0 & $-0.55 \phi^{\pm}_{2   }
+0.83\chi^{\pm}_{1 }$\\
Goldstone$^\pm_2$ & 0 &$0.31\tilde e_R^{\pm}-0.04 \phi^{\pm}_{1  }
+0.07\chi^{\pm}_{2 }+0.69\DE^\pm +0.65\delta^\pm $\\
\hline
\end{tabular}
\caption{Masses and compositions of the scalars and the Goldstone
bosons.
Parameters are chosen as follows $m_{\phi\chi }=m_{\DE\delta }=1$ TeV,
$\mu_2=100$ GeV, $h_\DE=0.6$, $A_\DE=1700$ GeV,
$\tan\beta=1.5$, $v_\delta /\sigma_R =1.5$,
$v_\DE/v_\delta =1.05$,
and $h_{\chi L}=0.1$.}
\end{table}

\begin{table}
\begin{tabular}{|l|r|l|} \hline
particle & mass [GeV] & composition \\ \hline
$H^0_1$ & 101 & $0.02\tilde\nu_R^{r}+0.02 \phi^{0r}_{1 }+1.00\chi^{0r}_{2 }
+0.03\delta^{0r} $\\
$H^0_2$ & 181 & $0.02 \phi^{0r}_{2  }+1.00\chi^{0r}_{1 } $\\
$H^0_3$ & 514 & $-0.01\tilde\nu_L^{r}-0.41 \tilde\nu_R^{r}
+0.03 \chi_2^{0r}-0.64\DE^{0r}
-0.65\delta^{0r} $\\
$H^0_4$ & 862 & $-0.90 \tilde\nu_R^{r}
+0.16\DE^{0r}+0.42\delta^{0r} $\\
$H^0_5$ & 1317 & $1.00\tilde\nu_L^{r}-0.09\phi^{0r}_{1  }
-0.01\delta^{0r} $\\
$H^0_6$ & 2252 & $-0.16 \tilde\nu_R^{r}
+0.02\chi^{0r}_{2 }+0.75\DE^{0r}-0.64\delta^{0r} $\\
$H^0_7$ & 7070 & $1.00 \phi^{0r}_{2  }-0.02\chi^{0r}_{1 } $\\
$H^0_8$ & 7099 & $-0.09\tilde\nu_L^{r}-1.00\phi^{0r}_{1   }
+0.02\chi^{0r}_{2 } $\\
$A_1$ & 182 & $-0.02 \phi^{0i}_{2   }+1.00\chi^{0i}_{1 } $\\
$A_2$ & 1317 & $1.00\tilde\nu_L^{i}+0.01 \tilde\nu_R^{i}-0.09\phi^{0i}_{1 }
$\\
$A_3$ & 1397 & $-0.01\tilde\nu_L^{i}+0.25 \tilde\nu_R^{i}
-0.62\DE^{0i} -0.74\delta^{0i}$\\
$A_4$ & 2869 & $0.01\tilde\nu_L^{i}-0.94 \tilde\nu_R^{i}
-0.33\DE^{0i} -0.04\delta^{0i}$\\
$A_5$ & 7070 & $1.00 \phi^{0i}_{2  }+0.02\chi^{0i}_{1 } $\\
$A_6$ & 7099 & $0.09\tilde\nu_L^{i}+1.00 \phi^{0i}_{1  }
+0.02\chi^{0i}_{2 }$\\
$H^\pm_1$ & 206 & $-0.02\tilde e_R^{\pm}-0.02 \phi^{\pm}_{1  }
+1.00\chi^{\pm}_{2 }-0.07\DE^\pm -0.03\delta^\pm $\\
$H^\pm_2$ & 1320 & $-1.00\tilde e_L^{\pm}+0.09 \phi^{\pm}_{2 }
$\\
$H^\pm_3$ & 1330 & $0.79\tilde e_R^{\pm}
+0.01\chi^{\pm}_{2 }+0.20\DE^\pm -0.58\delta^\pm $\\
$H^\pm_4$ & 1927 & $0.53\tilde e_R^{\pm}
-0.02\chi^{\pm}_{2 }-0.69\DE^\pm +0.48\delta^\pm $\\
$H^\pm_5$ & 7070 & $1.00 \phi^{\pm}_{1  }
+0.02\chi^{\pm}_{2 } $\\
$H^\pm_6$ & 7099 & $0.09\tilde e_L^{\pm}+1.00 \phi^{\pm}_{2  }
+0.02\chi^{\pm}_{1 }$\\
$H^{\pm\pm}_1$ & 225 & $0.69\DE^{\pm\pm} +0.73\delta^{\pm\pm} $\\
$H^{\pm\pm}_2$ & 1433 & $0.73\DE^{\pm\pm} -0.69\delta^{\pm\pm} $\\
Goldstone$^0_1$& 0 &$-0.02 \tilde\nu_R^{i}-0.02 \phi^{0i}_{1  }
+1.00\chi^{0i}_{2 }+0.07\DE^{0i} -0.06\delta^{0i}$\\
Goldstone$^0_2$ & 0 &$-0.22 \tilde\nu_R^{i}
+0.01\chi^{0i}_{2 }+0.71\DE^{0i} -0.67\delta^{0i}$\\
Goldstone$^\pm_1$ & 0 & $-0.02 \phi^{\pm}_{2   }
+1.00\chi^{\pm}_{1 }$\\
Goldstone$^\pm_2$ & 0 &$0.31\tilde e_R^{\pm}
+0.08\chi^{\pm}_{2 }+0.69\DE^\pm +0.65\delta^\pm $\\
\hline
\end{tabular}
\caption{Same as Table 1, except for $\tan\beta =50$.}
\end{table}

\input{epsf.sty}
\begin{figure}[t]
\leavevmode
\begin{center}
\mbox{\epsfxsize=15.cm\epsfysize=15.cm\epsffile{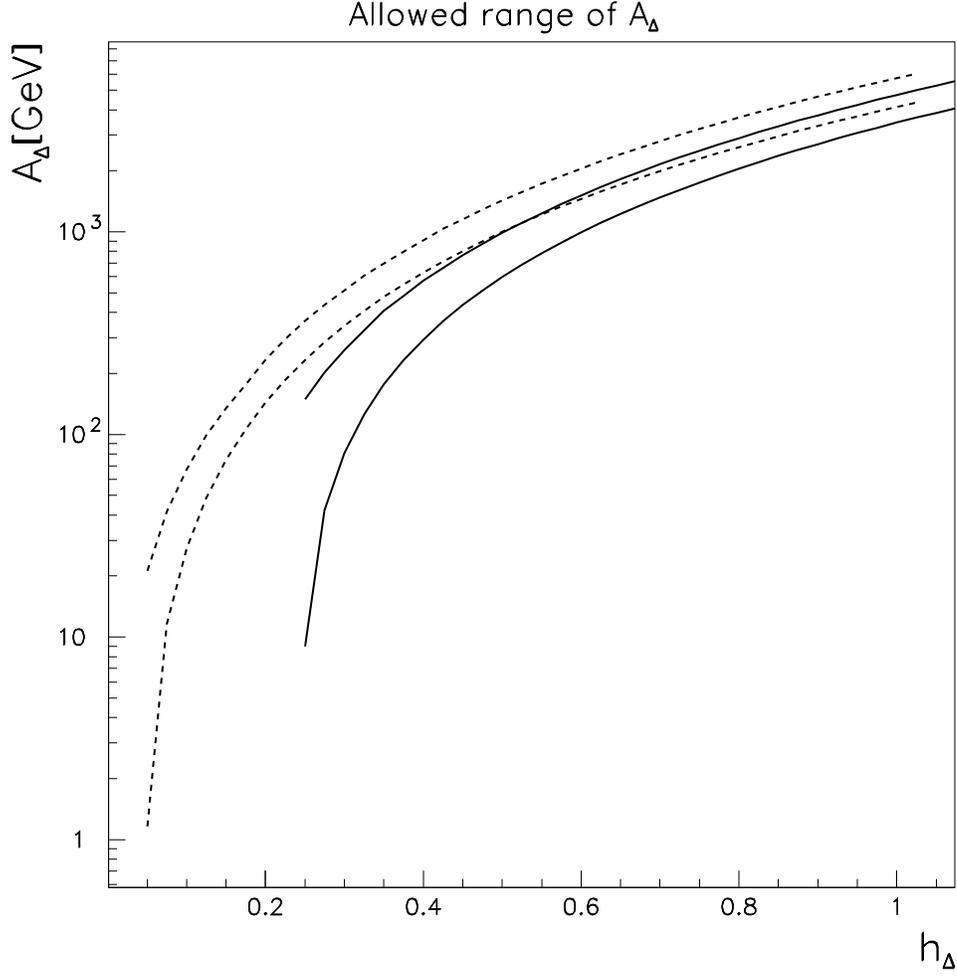}}
\end{center}
\caption{
The upper and lower limits of $A_\DE$ as a function of $h_\DE$.
The parameters are chosen as follows:
$m_{\phi\chi }=m_{\DE\delta }=1$ TeV,
$\tan\beta=50$, $v_\delta /\sigma_R =1.5$,
$v_\DE/v_\delta =1.05$,
and $h_{\chi L}=0.1$.
The solid curves correspond to $\mu_2=1$ TeV and the dashed curves
to $\mu_2=100$ GeV.}
\end{figure}

\begin{figure}[t]
\leavevmode
\begin{center}
\mbox{\epsfxsize=15.cm\epsfysize=15.cm\epsffile{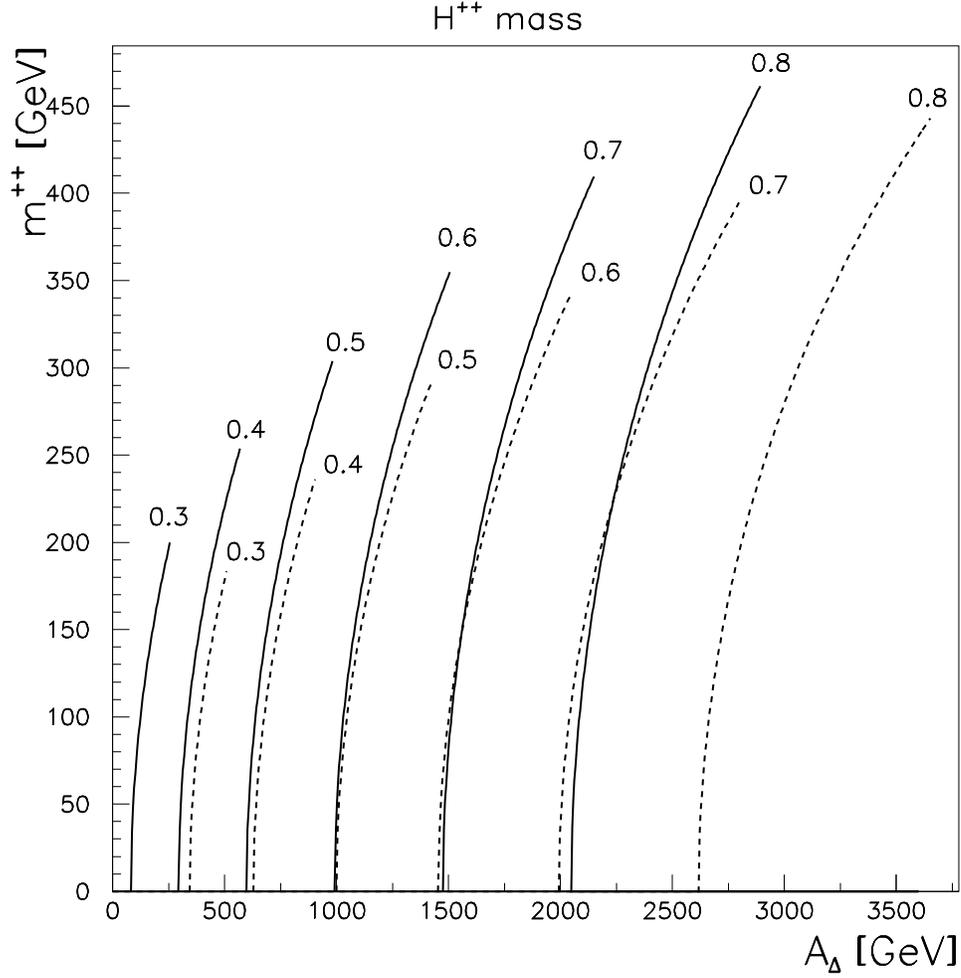}}
\end{center}
\caption{
The mass of the light doubly charged scalar as a function of the allowed
$A_\DE$ values for $h_\DE=0.3,\ldots,0.8$.
The parameters are chosen as follows:
$m_{\phi\chi }=m_{\DE\delta }=1$ TeV,
$\tan\beta=50$, $v_\delta /\sigma_R =1.5$,
$v_\DE/v_\delta =1.05$,
and $h_{\chi L}=0.1$.
The solid curves correspond to $\mu_2=1$ TeV and the dashed curves
to $\mu_2=100$ GeV.}
\label{fig2}
\end{figure}

\end{document}